# Corner states and topological transitions in two-dimensional higher-order topological sonic crystals with inversion symmetry

Zhan Xiong[1,#], Zhi-Kang Lin[1,#], Hai-Xiao Wang[1,2,#], Xiujuan Zhang[3], Ming-Hui Lu[3,4,†], Yan-Feng Chen[3,4], Jian-Hua Jiang[1,†]

[1]School of Physical Science and Technology, and Collaborative Innovation Center of Suzhou Nano Science and Technology, Soochow University, 1 Shizi Street, Suzhou, 215006, China

[1]School of Physical Science and Technology, Guangxi Normal University, Guilin, 541004, China

[2]National Laboratory of Solid State Microstructures and Department of Materials Science and Engineering, Nanjing University, Nanjing, China

[3]Collaborative Innovation Center of Advanced Microstructures, Nanjing University, Nanjing, China

[#]These authors contributed equally to this work.

[†]Correspondence and requests for materials should be addressed to jianhuajiang@suda.edu.cn, luminghui@nju.edu.cn

## ABSTRACT

Macroscopic two-dimensional sonic crystals with inversion symmetry are studied to reveal higher-order topological physics in classical wave systems. By tuning a single geometry parameter, the band topology of the bulk and the edges can be controlled simultaneously. The bulk band gap forms an acoustic analog of topological crystalline insulators with edge states which are gapped due to symmetry reduction on the edges. In the presence of mirror symmetry, the band topology of the edge states can be characterized by the Zak phase, illustrating the band topology in a hierarchy of dimensions, which is at the heart of higher-order topology. Moreover, the edge band gap can be closed without closing the bulk band gap, revealing an independent topological transition on the edges. The rich topological transitions in both bulk and edges can be well-described by the symmetry eigenvalues at the high-symmetry points in the bulk and surface Brillouin zones. We further analyze the higher-order topology in the shrunken sonic crystals where slightly different physics but richer corner and edge phenomena are revealed. In these systems, the rich, multidimensional topological transitions can be exploited for topological transfer among zero-, one- and two-dimensional acoustic modes by controlling the geometry.

## I. INTRODUCTION

Topological insulators are insulators with conducting boundary states that are topologically-protected and robust [1-2]. Unlike the conventional strong topological insulators where the gapless boundary states are protected by global symmetries such as time-reversal symmetry, topological crystalline insulators (TCIs) [3] are protected by crystalline symmetries that may not be preserved on the edge boundaries. TCIs are widely studied in electronic [4-7] and classical-wave (e.g., photonic [8-33] and acoustic [34-41]) systems with or without gapless edge states. Systems with gapped edge states due to edge boundary symmetry reduction may not support robust transport on the edges. However, though the edge states may open spectral gaps, the boundaries of edge boundaries may support topologically-protected hinge or corner states, yielding the concept of higher-order topological insulators (HOTIs) [42-72]. Generally, a $l$-th order topological insulator in $n$-dimension has topological boundary states on the $n-l$ dimesional boundaries. For instance, in two-dimensional (2D) systems, second-order topological insulators exhibit gapped edge states and in-gap corner states which emerge as the zero-dimensional (0D) topological boundary states of the gapped one-dimensional (1D) edge states. In the classical-wave analogs, such as acoustic band gap materials, higher-order topology provides a pathway for simultaneous wave trapping in the 1D edges and the 0D corners in a single system, which thus offers an unconventional approach for the manipulation of wave propagation and dynamics in artificial materials.

In this work, we show that 2D air-borne sonic crystals (SCs) with $C_2$-symmetry and controllable geometry offers an ideal platform for the study of higher-order topological physics. While the experimental demonstration has been revealed in Ref. [58], the main purpose of this work is to elaborate on the underlying physics and the topological band theory. The topological indices for the bulk and edges are elucidated here by analyzing the symmetry eigenvalues at the high-symmetry points in the bulk and surface Brillouin zones. The bulk-edge-corner correspondence are revealed using the first-principle calculation and the topological band theory. The rich topological transitions in the bulk and edges are illustrated where topological transfer between corner, edge and bulk states are demonstrated. In addition, we also study the higher-order topology for the shrunken $C_2$-symmetric SCs where similar physics but more complicated edge and corner phenomena are discovered.

This paper is organized as follows. In Sec. II, we introduce the $C_2$-symmetric SCs and their bulk band properties. In Sec. III, we study the higher-order topological physics and multi-dimensional topological transitions. In Sec. IV, we introduce the shrunken SCs and their bulk band properties. In Sec. V, the higher-order topology and topological transitions are elaborated for the shrunken SCs. We conclude and discuss potential future developments in Sec. VI.

## II. $C_2$-SYMMETRIC SONIC CRYSTALS AND BULK TOPOLOGICAL INDICES

The 2D square-lattice SCs studied here is depicted in Fig. 1(a). Each unit-cell consists of four block scatterers (indicated by the dark-blue blocks) made of epoxy with identical geometry (i.e., with length $l = 0.4a$ and width $w = 0.1a$, where $a$ is the lattice constant). The tunable geometry parameter is the rotation angle $\theta$, which varies from $-90°$ to $90°$ (note that $180°$ rotation does not change the unit-cell structure). The four meta-atoms are arranged to preserve the following glide symmetries, $G_x: (x,y) \rightarrow (x + \frac{a}{2}, \frac{a}{2} - y)$ and $G_y: (x,y) \rightarrow (\frac{a}{2} - x, y + \frac{a}{2})$. It should be noted that the unit-cell is not a primitive unit-cell, but the minimal unit-cell for the edges and corners studied in this work. The glide symmetries are not essential in the primitive unit-cell, but can be helpful in understanding the properties of the bulk and edge properties in the enlarged unit-cell. The unit-cell also has the following mirror symmetries, $M_x: (x,y) \rightarrow (-x, y)$ and $M_y: (x,y) \rightarrow (x, -y)$, as well as the two-fold rotation (i.e., inversion) symmetry, $C_2: (x,y) \rightarrow (-x, -y)$. The latter defines the parity of the Bloch wavefunctions at the high-symmetry points (HSPs) of the Brillouin zone and is the protective symmetry for the higher-order band topology in the SCs studied here.

The band gap between the second and third acoustic bands can be controlled by the rotation angle, $\theta$. The acoustic band gap vanishes when $\theta$ is equal to an integer of $90°$, whereas for other rotation angles a finite band gap can be obtained [see Fig. 1(b)]. The evolution of the acoustic bands at the M point is shown in Fig. 1(c). The band gap experiences a process of closing and reopening by rotating the four meta-atoms. During this process, the parity of the first two bands at the M point is reversed, indicating a topological transition. Specifically, for $-90° < \theta < 0°$, the first two bands have even parity at the M point, whereas for $0° < \theta < 90°$, the first two bands have odd parity at the M point.

We utilize the symmetry eigenvalues at the HSPs to characterize the bulk band topology. Following Ref. [34], the topological crystalline index can be expressed by the full set of the $\hat{C}_2$ eigenvalues at the HSPs. For a HSP denoted by the symbol $\Pi$, the $\hat{C}_2$ eigenvalues can only be $\Pi_p = e^{2\pi i(p-1)/2}$ with $p = 1$ (even-parity) or 2 (odd-parity). The HSPs of the Brillouin zone for 2D square lattices are $\Gamma, X, Y$ and $M$. The full set of the $\hat{C}_2$ eigenvalues at the HSPs is redundant due to time-reversal symmetry and the conservation of the number of bands below the band gap [72]. The minimum set of indices that describe the band topology can be obtained by using the following quantities [72]:

$$[\Pi_p] = \#\Pi_p - \#\Gamma_p, \tag{1}$$

where $\#\Pi_p$ and $\#\Gamma_p$ are the number of bands below the band gap with eigenvalues $\Pi_p$ and $\Gamma_p$, respectively, with $\Pi = X, Y$ and $M$. In this scheme, the symmetry eigenvalues at the $\Gamma$ point are taken as references. Since trivial atomic insulators the same symmetry eigenvalues for all HSPs, they have $[\Pi_p] = 0$ for all HSPs. In contrast, nonzero $[\Pi_p]$ indicates a band gap that is adiabatically disconnected with the trivial atomic insulator. That is, this band gap cannot be transformed into the trivial atomic insulator phase without breaking the protective symmetry or closing the band gap.

For the $C_2$-symmetric acoustic TCIs, the topological indices consist of three independent components which can be written in a compact form as

$$\chi = ([X_1], [Y_1], [M_1]). \tag{2}$$

From the $\hat{C}_2$ eigenvalues at the HSPs indicated in Fig. 1(b), we find that the topological index for the SCs with rotation angles $-90° < \theta < 0°$ is $\chi_1 = (-1, -1, 0)$ (we denote this class as TCI$_\alpha$). In contrast, for the SCs with rotation angles $0° < \theta < 90°$, the topological index is $\chi_2 = (-1, -1, -2)$ (we denote this class as TCI$_\beta$). The distinct topological indices indicate two topological distinct band gaps, which cannot be transformed into one another adiabatically without breaking the $\hat{C}_2$ symmetry.

In addition, the Wannier center distribution, which can be deduced from the bulk topological indices using the theory in Ref. [72], is different for the two types of TCIs. As shown in the insets of Fig. 1(c) (at the upper-left and upper-right corners), for the TCI$_\alpha$, the Wannier centers for the two bands below the band gap are located at the unit-cell center and the unit-cell corner, respectively. In contrast, for TCI$_\beta$, the two Wannier centers are located at the middle of the edges of a unit-cell. The Wannier center distributions reflect that the 2D bulk polarization of the two types of TCIs are the same, which is equal to $\boldsymbol{P} = (\frac{1}{2}, \frac{1}{2})$. This observation indicates that the underlying mechanisms for the edge and corner states in our systems are different from the mechanisms in Refs. [42, 45, 49, 51-53, 59-61, 68, 69] as the quadrupole topology or the mechanisms in Refs. [47, 50, 62-65] as the bulk polarizations.

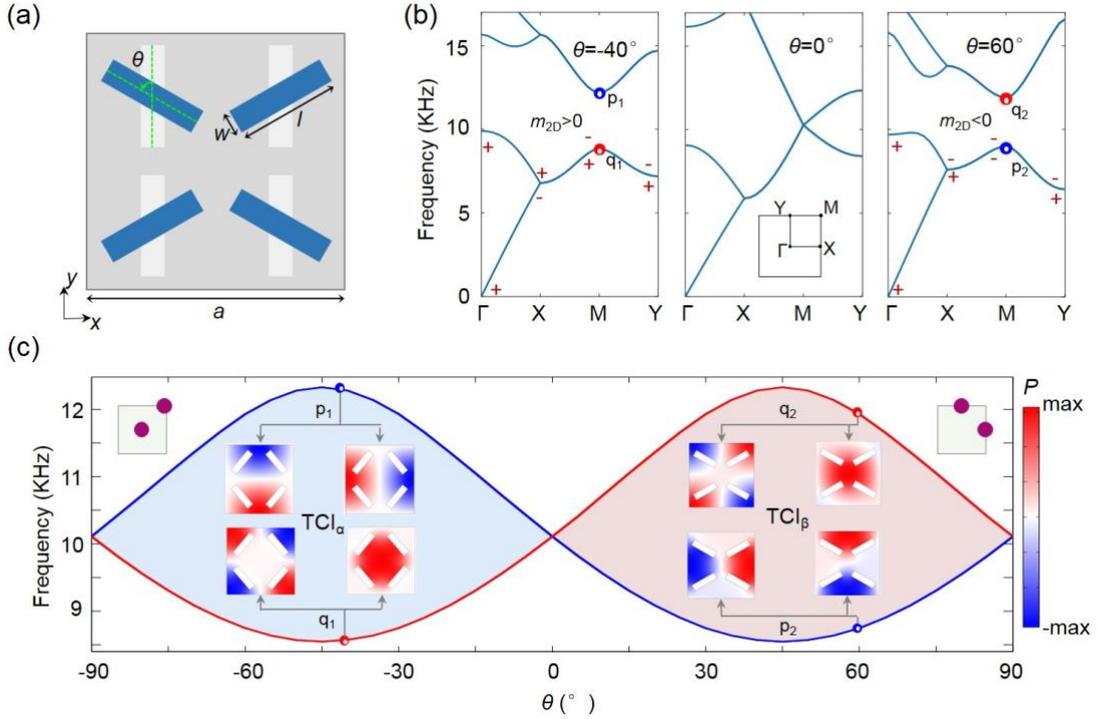

Fig. 1. (a) Schematic of the unit-cell of the air-borne SC. The four meta-atoms (represented by the blue blocks) are made of epoxy with a tunable rotation angle $\theta$. (b) Acoustic bands and parity eigenvalues (+ and −, for even and odd parity, respectively) of the $C_2$-symmetric SCs with different rotation angles $\theta = -40°$, $0°$ and $60°$. The red and blue dots represent the even and odd parity

eigenstates at the $M$ point, respectively. (c) The band edge frequencies at the $M$ point as functions of the rotation angle $\theta$. Middle-left and middle-right insets are the acoustic pressure profiles in a unit-cell for $\theta = -40°$ and $60°$, respectively. The upper-left and upper-right insets also show the Wannier centers (purple dots) for the two distinct topological phases, $TCI_\alpha$ and $TCI_\beta$.

### III. 2D MASSIVE DIRAC HAMILTONIAN FROM THE $k \cdot p$ THEORY

From the $k \cdot p$ theory, we find that the two classes of TCIs correspond to Dirac Hamiltonians with opposite Dirac masses, $m_{2D}$, as indicated in Fig. 1(b). Several basic elements should be pointed out before we going into details. First, to construct the $4 \times 4$ Dirac Hamiltonian, four acoustic Bloch bands are needed. Moreover, these four bands should include two bands below the band gap with identical parity and the other two bands above the band gap which have the parity opposite to the parity of the bands below the band gap. Here, the doubly degenerate bands below (above) the band gap are of the same parity, as shown in Figs. 1(b) and 1(c). On the other hand, the parities of the bands above and below the band gap are indeed opposite. Specifically, the first two bands below the band gap comprise a doublet with even (odd) parity for the rotation angles $-90° < \theta < 0°$ ($0° < \theta < 90°$), whereas the third and fourth bands above the band gap comprise a doublet with odd (even) parity. The acoustic pressure profiles (i.e., acoustic wavefunctions) of the four Bloch states at the M point are marked by the red (odd-parity; donated as $q$) and blue (even-parity; donated as $p$) dots in Figs. 1(b) and 1(c).

The double degeneracy on the Brillouin zone boundary lines $XM$ and $YM$ can be understood via the glide symmetries $G_x$ and $G_y$ [58] or the band folding picture [66]. Here we adopt the glide-symmetry approach. Combining the glide operators $G_j$ ($j = x, y$) and the time-reversal operator $T$, the anti-unitary operators $\Theta_j = G_j * T$, enable the double degeneracy at the Brillouin zone boundary because of the following logic: first, $\Theta_x^2 \psi_{n,k} = e^{ik_x a} \psi_{n,k}$, where $\psi_{n,k}$ is a Bloch wavefunction for the acoustic pressure field, and $n$ and $k$ are the band index and wavevector, respectively. At the Brillouin zone boundary $k_x a = \pi$, one finds $\Theta_x^2 = -1$ at the $XM$ line and thus induces fermion-like Kramers double degeneracy for all acoustic bands. Similarly, at the $YM$ line, i.e., $k_y a = \pi$, one finds $\Theta_y^2 = -1$ and hence fermion-like Kramers double degeneracy for all acoustic bands.

We find that the even-parity doublet consists of the $s$- and $d_{xy}$-like states, while the odd-parity doublet consists of the $p_x$- and $p_y$-like states [see the insets of Fig. 1(c)]. The degenerate $p_x$- and $p_y$-like ($s$- and $d$-like) eigenstates can form a new basis with finite orbital angular momenta (OAM), i.e., $p_\pm = p_x \pm ip_y$ ($d_\pm = s \pm id_{xy}$). In this way, the acoustic bands can be compared to the Bernevig-Hughes-Zhang model for the quantum spin Hall insulators where the "spins" are emulated by acoustic OAM. The different parities of the bands above and below the band gap ensures that the $\boldsymbol{k} \cdot \boldsymbol{p}$ Hamiltonian contains finite $\boldsymbol{k}$-linear couplings between the valence and conduction bands. This property is crucial for the simulation of massive Dirac Hamiltonian in acoustic systems in the aim of mimicking the quantum spin Hall effect. The exchange of the frequency order of the two doublets at the M point leads to the underlying physics similar to the "parity inversion" in the quantum spin Hall insulators. Therefore, there are two topologically distinct phases: the gapped phase with $m_{2D} > 0$ and the gapped phase with $m_{2D} < 0$ [see Fig. 1(b)], which have opposite parity order at the M point. These two cases correspond to, respectively, the TCI$_\alpha$ and TCI$_\beta$ defined in the previous section.

Explicitly, we consider the acoustic waves propagation in a 2D SC, obeying the following time harmonic wave equation

$$\kappa(\mathbf{r})\nabla \cdot [\rho^{-1}(\mathbf{r})\nabla\psi(\mathbf{r})] + \omega^2 \psi(\mathbf{r}) = 0, \tag{3}$$

where $\psi(\mathbf{r})$ represents the pressure field, and, $\kappa(\mathbf{r})$ and $\rho(\mathbf{r})$ denote the constitutive compress modulus and the mass density which depend on the 2D coordinate vector $\mathbf{r}$, respectively. Given the planar periodicity of the 2D SC in the *x-y* plane, solving Eq. (3) is equivalent to solving the following eigen-value problem,

$$H\psi_{n,\mathbf{k}}(\mathbf{r}) = \omega_{n,\mathbf{k}}^2 \kappa^{-1}(\mathbf{r})\psi_{n,\mathbf{k}}(\mathbf{r}), \tag{4}$$

where $H = -\nabla \cdot [\rho^{-1}(\mathbf{r})\nabla]$ is the "Hamiltonian". Although it does not have the dimension of energy, it is an Hermitian operator. Here, $\psi_{n,\mathbf{k}}(\mathbf{r})$ and $\omega_{n,\mathbf{k}}$ are, respectively, the wavefunction (for the pressure field) and the eigen-frequency of the acoustic Bloch wave with wavevector **k** in the *n*th band. The eigen-states are orthogonal, satisfying the normalization conditions $\delta_{nn'} = \int_{u.c.} \psi_{n,\mathbf{k}}^*(\mathbf{r}) \kappa^{-1}(\mathbf{r}) \psi_{n',\mathbf{k}}(\mathbf{r}) d\mathbf{r}$ where *u.c.* denotes the unit-cell, meaning that the integration is performed within a unit cell.

In the $\mathbf{k} \cdot \mathbf{p}$ theory, $\psi_{n,\mathbf{k}}(\mathbf{r})$ is expanded by the Bloch wavefunctions at the M point, i.e., $\psi_{n,\mathbf{k}}(\mathbf{r}) = e^{i\mathbf{q}\cdot\mathbf{r}} \sum_{n'} C_{n,n'} \psi_{n',\mathbf{K}}(\mathbf{r})$ where $\mathbf{q} = \mathbf{k} - \mathbf{K}$ with $\mathbf{K} = (\frac{\pi}{a}, \frac{\pi}{a})$ and $C_{n,n'}$ are the expansion coefficients. The wavevector difference $\mathbf{q}$ is treated as a small quantity and the Hamiltonian is expressed as the Taylor expansion of $\mathbf{q}$. Direct calculation yields the following $\mathbf{k} \cdot \mathbf{p}$ Hamiltonian,

$$H_{nn'} = \omega_n^2 \delta_{nn'} + \mathbf{q} \cdot \mathbf{P}_{nn'} + \cdots, \tag{5}$$

with $\omega_n$ the eigen-frequency of the *n*th band at the DP. Here, we keep only up to the linear terms of $\mathbf{q}$, while the higher-order terms are omitted. The matrix element of $\mathbf{P}_{nn'}$ is given by

$$\mathbf{P}_{nn'} = \int_{u.c.} \psi_{n,\mathbf{K}}^*(\mathbf{r}) \{-i[2\rho^{-1}(\mathbf{r})\nabla + \nabla\rho^{-1}(\mathbf{r})]\} \psi_{n',\mathbf{K}}(\mathbf{r}) d\mathbf{r}. \tag{6}$$

A crucial fact is that $\mathbf{P}_{nn'}$ is nonzero only when the *n* and *n'* bands are of different parities. In our system, there are four Bloch states interacting with each other. According to the field patterns shown in Fig. 1(c), there are two *p*-like states of odd parity, which we denote as $|p_x>$ and $|p_y>$. The other two Bloch states are of even parity, which are labeled as $|s>$ and $|d_{xy}>$. According to the mirror symmetries of these four states, the $\mathbf{q}$-linear term in the basis of $(|s>, |d_{xy}>, |p_x>, |p_y>)^T$ is

$$\mathbf{q} \cdot \mathbf{P} = \begin{pmatrix} 0 & 0 & aq_x & bq_y \\ 0 & 0 & cq_y & dq_x \\ a^*q_x & b^*q_y & 0 & 0 \\ c^*q_y & d^*q_x & 0 & 0 \end{pmatrix}. \tag{7}$$

Since the other acoustic bands are far away from the four Bloch bands considered here at the M point, the $\mathbf{k} \cdot \mathbf{p}$ theory can be restricted to the Hilbert space consisting of only the aforementioned four Bloch states.

We now study the constraints on the Hamiltonian due to the glide symmetries. The glide operation $G_y$ transforms the four states as follows: $\langle p_y|G_y|p_x\rangle = 1$, $\langle p_x|G_y|p_y\rangle = -1$, $\langle d_{xy}|G_y|s\rangle = 1$, and $\langle s|G_y|d_{xy}\rangle = -1$ for the $G_y$ operation.

Similarly, for the $G_x$ operation, $\langle p_y|G_x|p_x\rangle = -1$, $\langle p_x|G_x|p_y\rangle = 1$, $\langle d_{xy}|G_x|s\rangle = 1$, and $\langle s|G_x|d_{xy}\rangle = -1$. The invariance of the Hamiltonian under $\Theta_y$ implies

$$\Theta_y H(\mathbf{q})\Theta_y^{-1} = H(\Theta_y \mathbf{q}). \tag{8}$$

Note that $\Theta_y \mathbf{q} = G_y T \mathbf{q} = -G_y \mathbf{q} = (q_x, -q_y)$. When acting on the Hamitonian, the time-reversal operator $T$ is explicityly the complex conjugation. Using Eq. (8), we find that

$$a = d^*, \quad b = c^*. \tag{9}$$

As a consequence, the $\mathbf{k}\cdot\mathbf{p}$ Hamiltonian becomes

$$H(\mathbf{q}) = \begin{pmatrix} \omega_s^2 & 0 & aq_x & bq_y \\ 0 & \omega_d^2 & b^*q_y & a^*q_x \\ a^*q_x & b^*q_y & \omega_{p_x}^2 & 0 \\ bq_y & aq_x & 0 & \omega_{p_y}^2 \end{pmatrix}. \tag{10}$$

The doubly degenerate states can be reorganized into the "spin-up" states, $|d_+>=\frac{1}{\sqrt{2}}(|s>+i|d_{xy}>)$ and $|p_+>=\frac{1}{\sqrt{2}}(|p_x>+i|p_y>)$, and the "spin-down" states, $|d_->=\frac{1}{\sqrt{2}}(|s>-i|d_{xy}>)$ and $|p_->=\frac{1}{\sqrt{2}}(|p_x>-i|p_y>)$. Then, the Hamiltonian can be rewritten in the basis of $(|d_+>,|p_+>,|d_->,|p_->)^T$ as,

$$H_{Dirac} = \begin{pmatrix} M_+ & N \\ N^\dagger & M_- \end{pmatrix}, \tag{11}$$

where $M_+$ and $M_-$ refer to the spin-up and spin-down Blocks, respectively, which are

$$M_\pm = \begin{pmatrix} \omega_d^2 & \frac{1}{2}[(a+a^*)q_x \pm i(b^*-b)q_y] \\ \frac{1}{2}[(a+a^*)q_x \pm i(b-b^*)q_y] & \omega_p^2 \end{pmatrix}. \tag{12}$$

and, $N$ is inter-spin coupling matrix,

$$N = \begin{pmatrix} 0 & \frac{1}{2}[(a-a^*)q_x + i(b+b^*)q_y] \\ \frac{1}{2}[(a^*-a)q_x + i(b^*+b)q_y] & 0 \end{pmatrix}. \tag{13}$$

Eq. (11) is a 2D massive Dirac Hamiltonian where the Dirac mass is given by,

$$m_{2D} = \omega_d^2 - \omega_p^2. \tag{14}$$

Therefore, the band inversion between the odd- and even-parity bands correspond to the change of the 2D Dirac mass, indicating a topological transition similar to the normal insulator ($m_{2D} > 0$) to quantum spin Hall insulator ($m_{2D} < 0$) transition. Therefore, the $\boldsymbol{k} \cdot \boldsymbol{p}$ theory reveals that our SCs simulate the quantum spin Hall effect in acoustic systems where the "spins" are emulated by OAM.

### IV. EDGE STATES AT THE BOUNDARIES BETWEEN TCI$_\alpha$ AND TCI$_\beta$

Considering the cases with edge boundaries along the $x$ and $y$ directions between the two SCs with different rotation angles, $\theta_1 = -25°$ and $\theta_2 = 55°$. As shown in Figs. 2(a) and 2(b), the edge physics are similar to that of magnetized quantum spin Hall insulators where the broken time-reversal symmetry leads to gapped edge spectrum [73]. Away from the time-reversal invariant wavevectors, $k_{x/y} = 0, \pi/a$, the edge states exhibit finite OAM (i.e., "spin"-polarizations), as indicated by the red and blue arrows in Figs. 2(a) and 2(b). The finite OAM are manifested in the phase vortices of the acoustic edge wavefunctions as well as the winding of the acoustic Poynting vectors [the green arrows; see Figs. 2(c) and 2(d)]. At the time-reversal invariant wavevectors, $k_{x/y} = 0, \pi/a$, the OAM vanishes due to the time-reversal symmetry. The edge states become the eigenstates of the mirror symmetry of the edge structure. Specifically, the edge along the $x$ ($y$) direction has the mirror symmetry $M_x$ ($M_y$). The mirror eigenvalues of the edge states are labeled in Figs. 2(a) and 2(b) ("even" and "odd" standing for the mirror eigenvalues $1$ and $-1$, respectively).

The gapped edge states, which is quite common for TCIs, can also be understood via the glide symmetries. Since the glide symmetries are broken on the boundaries, the double degeneracy at the $k_y = \frac{\pi}{a}$ ($k_x = \frac{\pi}{a}$) point is no longer guaranteed by the glide symmetries. Some special configurations, where the glide symmetries are restored on the edges, are shown later [see Sec. VI]. Except for these special cases, the gapped edge states can be described by the 1D massive Dirac equations,

$$H_j = v_j(k_j - \frac{\pi}{a})\sigma_z + m_j\sigma_y \tag{15}$$

with $j = x, y$. Here, the spin-up and spin-down states are the edge states with positive and negative OAM along the $z$ direction, respectively [see Figs. 2(c) and 2(d)], while the eigenstates of the $\sigma_y$ operator are the even and odd parity edge states at the $k_{x/y} = \frac{\pi}{a}$. The Dirac masses $m_x$ and $m_y$ are related, respectively, to the size of the edge band gap for the edge along the $x$ and $y$ directions. Explicitly,

$$m_x = \frac{1}{2}(\omega_{odd}^{edgex} - \omega_{even}^{edgex}), \qquad (16)$$

$$m_y = \frac{1}{2}(\omega_{odd}^{edgey} - \omega_{even}^{edgey}), \qquad (17)$$

where $\omega_{odd}^{edgex}$ and $\omega_{even}^{edgex}$ ($\omega_{odd}^{edgey}$ and $\omega_{even}^{edgey}$) denote the frequencies of the odd and even edge states at $k_x = \frac{\pi}{a}$ ($k_y = \frac{\pi}{a}$), respectively. Importantly, the mirror symmetries lead to quantized Zak phases for the edge states. The edge states can be trivial or topological 1D insulators. The nontrivial Zak phase is indicated by the parity inversion for the edge band below the edge band gap, i.e., negative Dirac mass. Thus, the edge band gap for the edge along the $x$ direction has nontrivial topology (i.e., Zak phase $\pi$), whereas the edge band gap for the edge along the $y$ direction has trivial Zak phase (i.e., Zak phase 0). In other words, the polarization of the band gap for the edge along the $x$ direction is $p_x^{edgex} = \frac{1}{2}$, whereas the polarization of the band gap for the edge along the $y$ direction is $p_y^{edgey} = 0$.

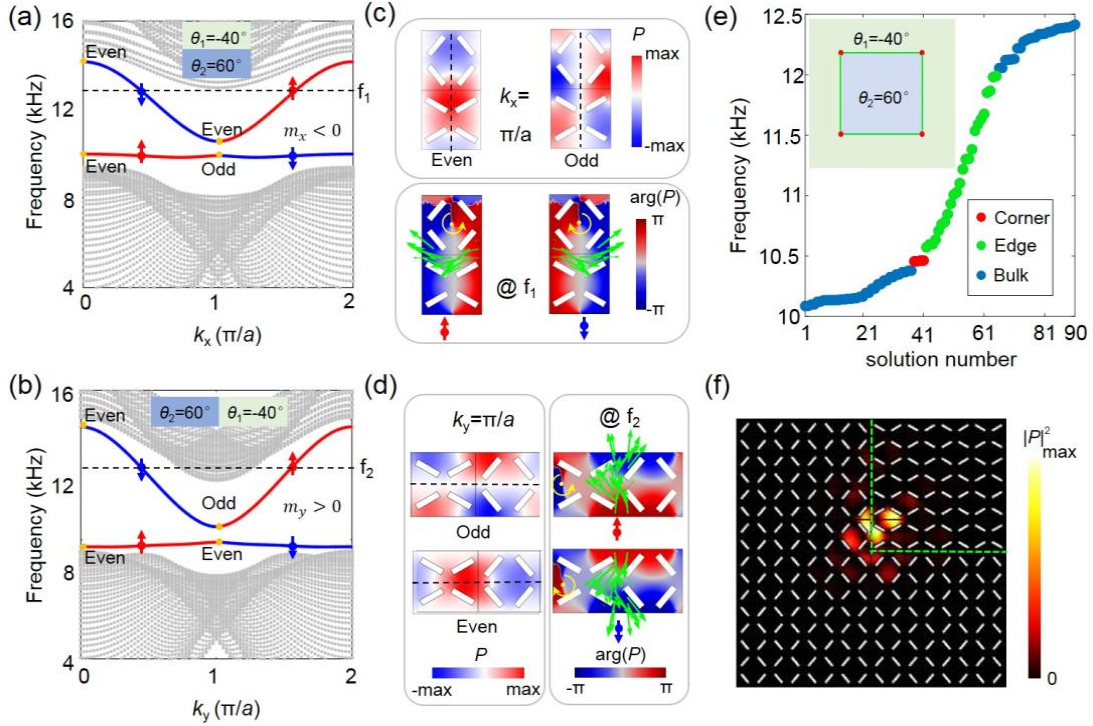

Fig. 2. (a)-(b) Calculated spectra of the edge boundaries between the SC with $\theta = -40°$ and the SC with $\theta = 60°$ for the edges along the $x$ (a) and $y$ (b) directions, respectively. (c)-(d) Wavefunctions of the edge states at $k_{x/y} = \pi/a$ and the edge states at frequencies $f_1$ and $f_2$. For $k_{x/y} = \pi/a$, the edge states are either even parity or odd parity (the dashed lines indicate the mirror plane). For the edge states at frequencies $f_1$ and $f_2$, the edge states are either pseudo-spin up or pseudo-spin down. The yellow arrows in the acoustic pressure phase profiles, $\arg(P)$, indicate the winding directions of the phase vortices (the yellow dots indicate the vortex centers). The green arrows indicate the momentum ("Poynting vector") distributions of the acoustic edge states. (e) Calculated spectrum for the finite structure (depicted in the inset) with both edge and corner boundaries between the $TCI_\alpha$ with $\theta = -40°$ and $TCI_\beta$ with $\theta = 60°$. The spectrum of the four degenerate corner states is indicated by the red dots. (f) Acoustic wavefunction (i.e., acoustic pressure profile) for one of the corner state. The green dashed lines indicate the edge boundaries.

## V. CORNER STATES

In Fig. 2(e) we show the eigenstates spectrum for a box-shaped structure where the $TCI_\beta$ is surrounded by the $TCI_\alpha$. The structure has both edge and corner boundaries. Thus, the eigenstates include bulk (blue dots), edge (green dots) and corner (red dots) states. There are four degenerate corner states emerging in the common spectral gap of the edge and bulk. Each corner has one corner state. The wavefunction of the corner state is shown in Fig. 2(f).

The emergence of the topological corner states can be understood via several pictures depicted in Fig. 3. First, the description of the edge states as 1D massive Dirac equations gives access to the understanding of corner states as the Jackiw-Rebbi solitons as illustrated in Fig. 3(a). For the corners connecting the edges along the $x$ and $y$ directions, the opposite Dirac masses for these edges lead to the emergence of the Jackiw-Rebbi solitons at the four corners [74].

Second, the corner charge can also be obtained from counting the edge polarizations, as illustrated in Fig. 3(b). In higher-order TCIs without quadrupole topology, as in the cases studied here, the corner charge is determined by the addition of the edge polarizations along the $x$ and $y$ directions [42]. That is, $q_{corner} = (p_x^{edgex} + p_y^{edgey})$ mod 1 where $p_x^{edgex}$ and $p_y^{edgey}$ denote the edge polarization along the $x$ and $y$ directions, respectively. From the analysis in the previous sections, we find that $q_{corner} = \frac{1}{2}$, which dictates nontrivial, topological corner states. This picture is consistent with the fact that the corner states can be understood as the boundary states between the topological 1D edge states along the $x$ direction (Zak phase [75] $\pi$) and the trivial 1D edge states along the $y$ direction (Zak phase 0).

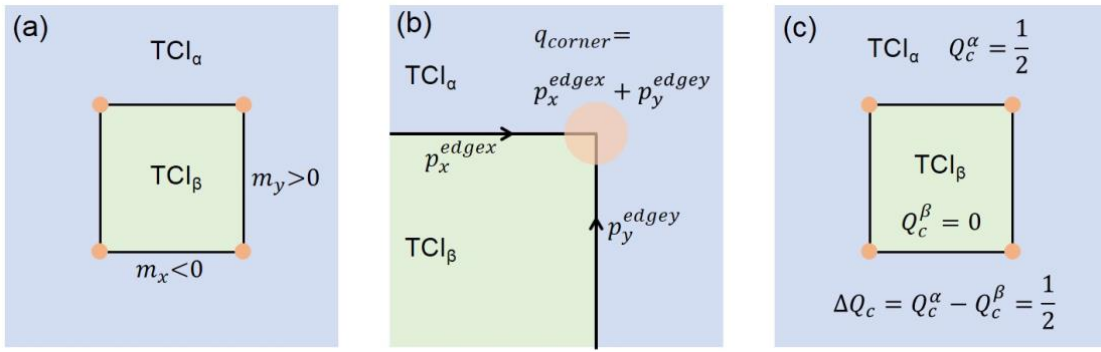

Fig. 3. Three different pictures to understand the emergence of the topological corner states. (a) Corner states emerging as the Jackiw-Rebbi solitons at the corner boundary between massive-Dirac edge states with opposite signs of Dirac mass, e.g., $m_y > 0$ and $m_x < 0$. (b) Emergence of corner states as indicated by the nontrivial corner charge as from counting the contributions from the edge polarizations, i.e., $p_x^{edgex} = \frac{1}{2}$ and $p_y^{edgey} = 0$, thus $q_{corner} = \frac{1}{2}$. (c) Emergence of corner states

due to nontrivial corner topological index. The corner topological index from the symmetry-representations for the TCI$_\alpha$ is $Q_c^\alpha = \frac{1}{2}$, whereas for the TCI$_\beta$ it is $Q_c^\beta = 0$. The nontrivial difference, $\Delta Q = Q_c^\alpha - Q_c^\beta$, indicates the emergence of the topological corner states.

Third, apart from the above edge analysis, the emergence of the corner states can be understood from the bulk topological invariants, as illustrated in Fig. 3(c). According to Ref. [72], the bulk topological invariants in Eq. (2) are connected to the bulk-induced corner topological index, $Q_c$, as

$$Q_c = \frac{1}{4}(-[X_1] - [Y_1] + [M_1]) \bmod 1. \qquad (18)$$

From the bulk topological indices $\chi = ([X_1], [Y_1], [M_1])$, we find that for TCI$_\alpha$ [i.e., $([X_1], [Y_1], [M_1]) = (-1, -1, 0)$] the corner topological index is $Q_c^\alpha = \frac{1}{2}$, whereas for TCI$_\beta$ [i.e., $([X_1], [Y_1], [M_1]) = (-1, -1, -2)$] the corner topological index is $Q_c^\beta = 0$. The difference between the two bulk-induced corner topological indices is, $\Delta Q_c = \left(Q_c^\alpha - Q_c^\beta\right) \bmod 1 = \frac{1}{2}$. The nontrivial difference of the corner topological indices of the two TCIs, $\Delta Q_c = \frac{1}{2}$, indicates the emergence of the corner states in the corner boundaries between the two TCIs. The bulk-edge-corner correspondence is revealed through topological transitions at the bulk and edges in the next section.

We now study the robustness of the corner states via numerical simulations. The results are presented in Fig. 4. Because of the localized nature of the corner states, they are naturally robust to disorder and perturbations away from the corner. We thus focus on the situations where the perturbations are near the corner of concern. Fig. 4 show that for three different cases, the frequency of the corner states changes little even strong deformations are introduced on the corner. In the meanwhile, counterintuitively, the wavefunctions of the corner states experience considerable changes. For all these cases with deformations, the corner state remain existing within the bulk band gap. These properties indicate the robustness of the topological corner modes in the higher-order topological SCs. Similar behaviors in the eigen-frequency and the wavefunction of the corner state have been found before in topological bound states on dislocations [28] and the topological surface states due to the Dirac points [24] in photonics.

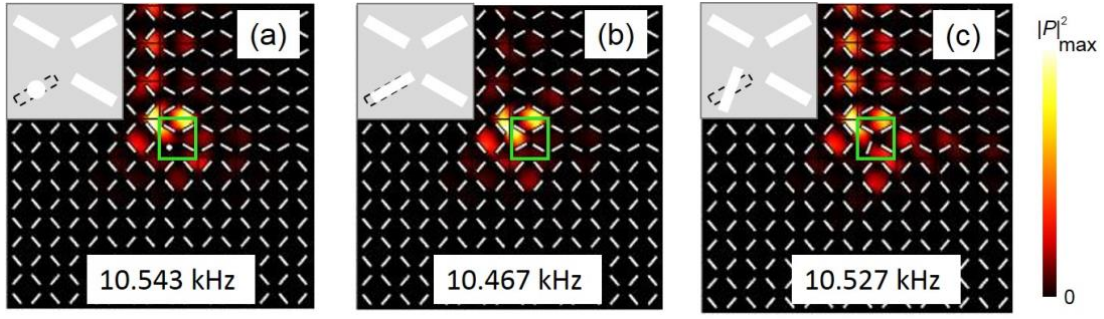

Fig. 4. Robustness of the corner state between the SC with $\theta = -40°$ and the SC with $\theta = 60°$. Wavefunctions (hot color) and frequencies (inset) of the corner state when the epoxy block (dashed) in the unit-cell near the corner is replaced by (a) an epoxy dot, (b) a shifted epoxy block, and (c) a rotated epoxy block. The inset at the upper-left show the structure of the deformed unit-cell.

## VI. TOPOLOGICAL TRANSITION IN A HIERARCHY OF DIMESIONS

Since the topological transition in the bulk is triggered by the rotation angle $\theta$, similar transitions may take place in the edges. In Fig. 5 we study the topological transitions in the edges by considering the $x$ and $y$ edge boundaries between the $TCI_\alpha$ with $\theta_1 = -40°$ and the $TCI_\beta$ with varying $\theta_2$. In Fig. 5(a) we plot the frequencies of the odd and even edge modes at $k_x = \frac{\pi}{a}$ for the edge along the $x$ direction, i.e., $\omega_{odd}^{edgex}$ and $\omega_{even}^{edgex}$, as functions of the rotation angle in the $TCI_\beta$, i.e., $\theta_2$. We find that the Dirac mass of the edge states along the $x$ direction, $m_x$, goes from negative to positive when the angle $\theta_2$ decreases from 90° to 0°. A topological transition takes place at $\theta_2 = 40°$ where $m_x$ vanishes. We find that such a topological transition is due to the emergence of glide symmetry at the edge, $G_{edge,x} := (x, y) \to (x + \frac{a}{2}, -y)$ (the origin of the coordinate is at the center of the edge boundary), as illustrated in Fig. 5(b). Due to such glide symmetry, the edge states along the $x$ direction become gapless, as shown in Fig. 5(c). The gap closing at the $k_x = \frac{\pi}{a}$ point is due to the following logic: Combining the glide operation and the time-reversal operation yields $\Theta_{edge,x} = G_{edge,x} T$. When acting on the edge wavefunctions $\varphi_{n,k_x}$ ($n$ is the edge band index, and $k_x$ is the wavevector), $\Theta_{edge,x}^2 \varphi_{n,k_x} = e^{ik_x a} \varphi_{n,k_x}$. Therefore, at the $k_x = \frac{\pi}{a}$ point, $\Theta_{edge,x}^2 \varphi_{n,k_x} = -\varphi_{n,k_x}$, which leads to the Kramers-like degeneracy for the edge states. The gap closing at $\theta_2 = 40°$ is hence protected by the emergent edge glide symmetry.

We find that, for general cases, the edge glide symmetry emerge when $\theta_2 = -\theta_1$, leading to edge gap closing and edge topological transition.

Similar scenario for the edge band gap closing and edge topological transition is found for the edge states along the $y$ direction at the same condition, $\theta_2 = -\theta_1$, see Figs. 5(d)-5(f). Besides, the edge band gap (for the edges along the $x$ and $y$ directions) closes at $\theta_2 = 0°$ and $90°$ due to the bulk band gap closing at these points, since the edge states emerge due to the bulk band topology. We hence find that in the higher-order SCs studied in this work, there are multiple topological transitions in different dimensions: In the bulk, there are topological transitions at $\theta = 0°$ and $90°$, etc. (i.e., the integer multiples of $90°$). Apart from the bulk topological transitions, there is a topological transition at $\theta_2 = -\theta_1$ (or equivalently, $180° - \theta_1$).

We emphasize that the topological transitions at the edges, interestingly, does not change the corner topological index $q_{corner}$ which is solely determined by the bulk. This property can be observed via the fact that the edge Dirac masses, $m_x$ and $m_y$, always have opposite signs, before and after the edge topological transition at $\theta_2 = -\theta_1$. Therefore, the corner topological index remains nontrivial, i.e., $q_{corner} = \frac{1}{2}$, across the edge topological transition. This observation indicates that the emergence of the corner states is essentially due to the bulk topological properties.

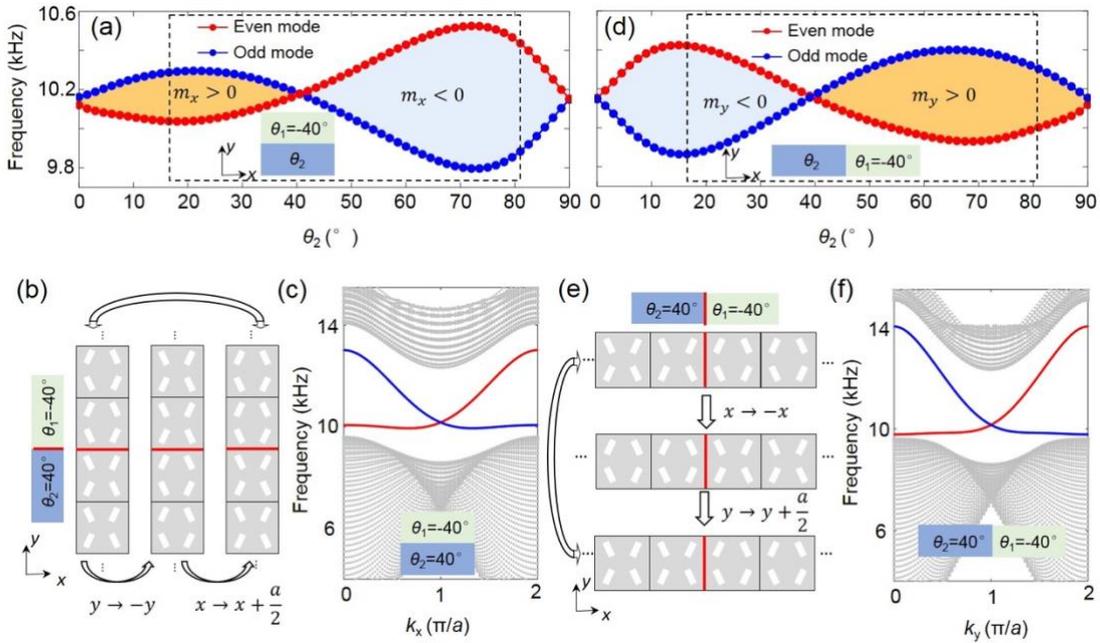

Fig. 5. Topological transitions in a hierarchy of dimensions when $\theta_1 = -40°$. (a) Topological "phase diagram" for the edge states propagating along the $x$ direction as represented by the frequencies of the odd and even modes at $k_x = \frac{\pi}{a}$. The transitions occur at the parity inversion (i.e., gap closing) points where the odd and even modes become degenerate. (b) Illustration of the glide symmetry at the edge boundary along the $x$ direction when $\theta_2 = -\theta_1$. (c) Gap-closing edge states along the $x$ direction for the case with $\theta_2 = -\theta_1 = 40°$. (d) Topological "phase diagram" for the edge states propagating along the $y$ direction as represented by the frequencies of the odd and even modes at $k_y = \frac{\pi}{a}$. (e) Illustration of the glide symmetry at the edge boundary along the $y$ direction when $\theta_2 = -\theta_1$. (c) Gap-closing edge states along the $y$ direction for the case with $\theta_2 = -\theta_1 = 40°$.

Nevertheless, the close of the edge band gaps diminishes the corner states, since the corner states have to be stabilized by the edge band gaps. Precisely speaking, the corner states emerge in the common spectral gap of the edge states along the x and y directions. We label the regions with a common spectral gap for the $x$ and $y$ edges by the dashed boxes in Fig. 5. In Fig. 6, we plot the evolution of the local density of states in the bulk, edge and corner regions [schematically illustrated in Fig. 6(a)] with the rotation angle $\theta_2$. As shown in Figs. 6(b) and 6(f), without the common spectral gap for the $x$ and $y$ edges, the corner resonances disappear (In fact, at these conditions, the corner states merge into the edge, as shown in Ref. [58]). In contrast, with the common spectral gap for the $x$ and $y$ edges, a clear corner resonance emerge in the edge spectral gap [Figs. 6(c), 6(d) and 6(e)]. For the sake of graphic presentations in Fig. 6, a finite Lorentzian broadening of 10 Hz is used in the calculation of the density of states. The rich topological transitions in the bulk and edges (Fig. 5) induced by controlling the rotation angles of the epoxy scatterers enable versatile control of the corner, edge and bulk states shown in Fig. 6, which has been demonstrated in the experiments in Ref. [58].

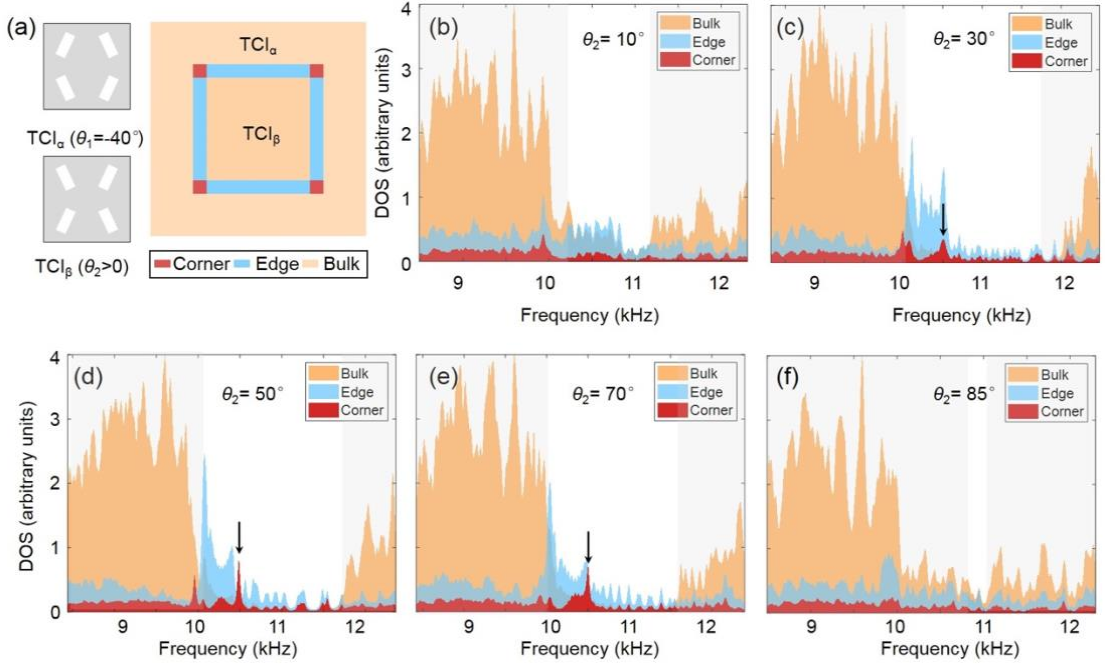

Fig. 6. (a) Illustration of the structure with edge and corner boundaries between $TCI_\alpha$ with $\theta_1 = -40°$ and $TCI_\beta$ with $0° < \theta_2 < 90°$. Different colors represent different regions for the calculation of local density of states (DOS). The DOS for the corner, edge and bulk are obtained by integration of local DOS over the corner, edge and bulk regions, respectively. There are four corner (edge) regions. Each corner region contains $2 \times 2$ unit-cells, while each edge region contain $2 \times 10$ unit-cells. (b)-(f) DOS for the corner, edge and bulk regions for $\theta_2 = 10°, 30°, 50°, 70°$ and $85°$, respectively. The corner resonances are indicated by the black arrows.

## VII. SHRUNKEN SONIC CRYSTALS AND BULK BAND PROPERTIES

We now study the higher-order topology and its boundary manifestations in SCs with certain deformations. Specifically, we consider SCs with the four epoxy blocks shrinking toward the unit-cell center by a distance $d = 0.02a$, as depicted in Fig. 7(a). The unit-cell shown in the figure is the primitive unit-cell which has the $C_2$ rotation symmetry. It is found that the underlying higher-order topological physics in the shrunken SCs is slightly different, and the edge and corner manifestations are richer. The shrunken SCs serve as good examples to confirm the interesting physics discovered in the original SCs studied in the previous sections. We emphasize that although the unit-cells of the shrunken SCs do not have glide symmetry, the mirror symmetries, $M_x$ and $M_y$, are preserved.

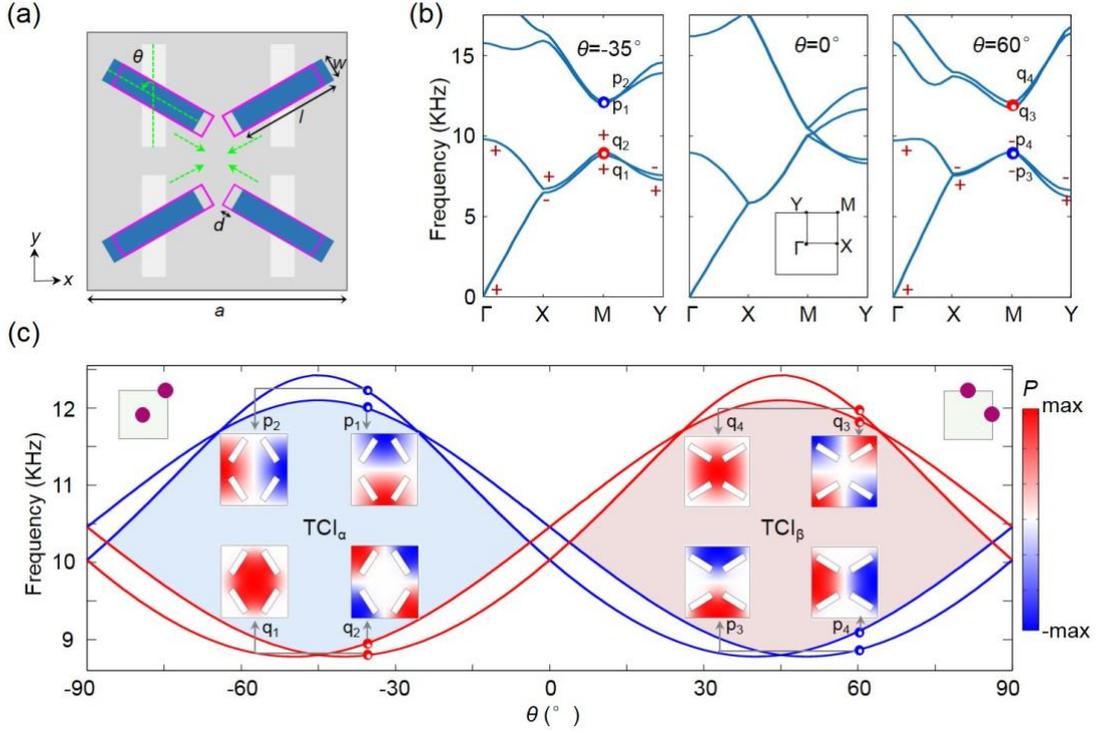

Fig. 7. Schematics of the unit-cell of the shrunken SCs. There are four epoxy blocks in each unit-cell. To illustrate the shrinking deformation, we show the original positions of the epoxy blocks by the blue regions, while the position after deformation is shown by the magenta-framed blocks. The shrinking distance is $d$, the tunable rotation angle is $\theta$. (b) Bulk energy bands of shrunken SCs with $\theta = -35°$, $0°$ and $60°$. The red symbols + and – denote, respectively, the eigenvalues 1 and -1 of the $C_2$ symmetry at the HSPs of the Brillouin zone. (c) Band edge frequencies of the first four bulk bands at the M point versus the rotation angle $\theta$. The insets in the blue and red regions indicate the acoustic pressure profiles at the M point for the cases with $\theta = -35°$ and $60°$, respectively.

There are several consequences due to the symmetry reduction. First, shrinking removes the glide symmetries, $G_x$ and $G_y$, in the shrunken SCs, leading to degeneracy splitting of the bulk bands at the M point [see Figs. 7(b) and 7(c)]. However, the $C_2$-symmetry eigenvalues for the first two bands do not change for most of the rotation angles (precisely, for the blue and red regions in Fig. 7(c)). Therefore, the topological properties of the concerned band gap (i.e., the band gap between the second and the third bands) remain the same as the original SCs. Nevertheless, the removal of the glide symmetries do change the topological phase diagram slightly, as shown in Fig. 7(c).

The regions with complete bulk band gap become narrower, because the regions, $-5° < \theta < 5°$ and $85° < \theta < 95°$, are now gapless. The acoustic band gap in the region $-85° < \theta < -5°$ correspond to the $TCI_\alpha$, while the band gap in the region $5° < \theta < 85°$ correspond to the $TCI_\beta$. The Wannier centers for these two distinct phases are shown in the upper-left and upper-right insets of Fig. 7(c).

## VIII. HIGHER-ORDER TOPOLOGY AND MULTI-DIMENSIONAL TOPOLOGICAL TRANSITIONS FOR SHRUNKEN SONIC CRYSTALS

### A. Edge states

Since the shrunken SCs have the same topological properties as the original SCs, their edge and corner physics must also be similar. In Figs. 8(a)-8(d), we show the dispersions and eigenstates properties of the edge and corner states. It is seen that the edge states resemble the gapped edge states with "spin" polarizations in the original SCs, where the "spins" are emulated by acoustic OAM. Like the edge properties of the original SCs, at the time-reversal invariant wavevectors, $k_{x/y} = 0, \frac{\pi}{a}$, the edge states are mirror eigenstates. Their mirror eigenvalues are labeled in the figure (even for 1, and odd for -1). These edge properties are the same as what we have found for the original SCs in previous sections. In accordance, the corner states emerge at the corner boundary between $TCI_\alpha$ and $TCI_\beta$, as shown in Figs. 8(e) and 8(f). The edge Hamiltonian can still be described by Eq. (15).

The robustness of the corner states in shrunken SCs is studied via numerical simulations. The results presented in Fig. 9 show that the existence of the corner states is robust against disorders and strong deformations even when they are close to the corner. In addition, the change in the frequency of the corner state is small, although its wavefunction is considerably modified by the disorder.

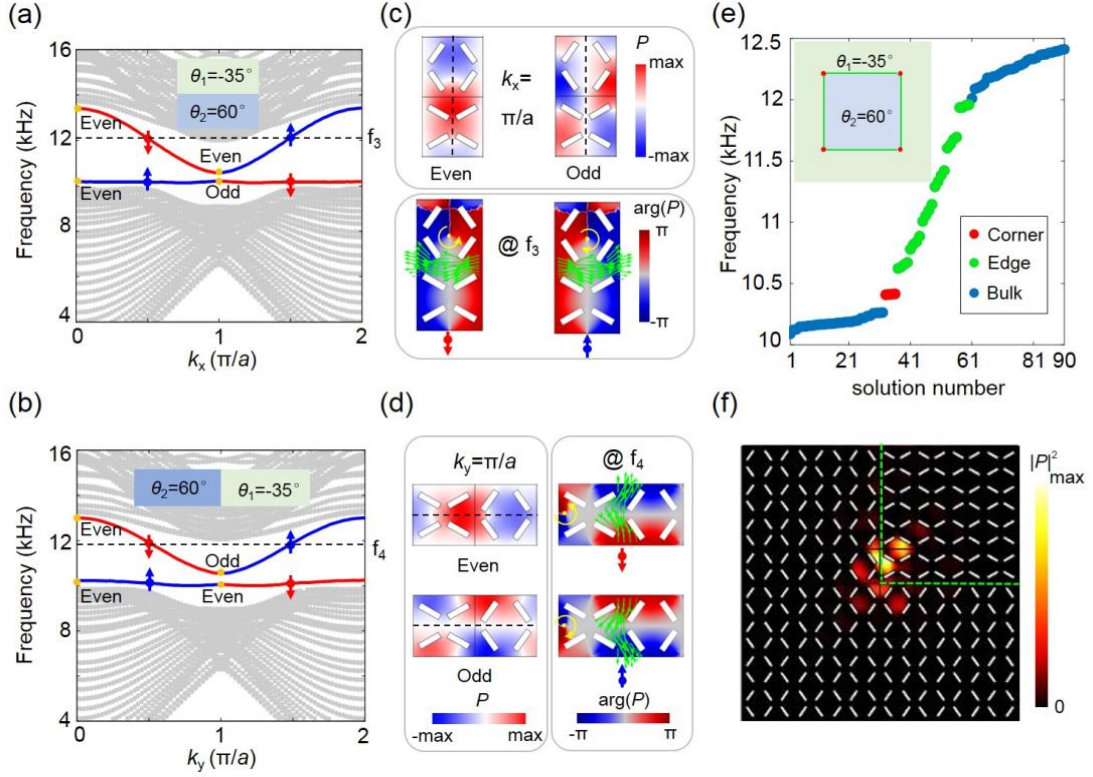

FIG. 8. (a)-(b) Simulated spectra of the edge boundaries between the shrunken SC with $\theta = -35°$ and the shrunken SC with $\theta = 60°$ for the edges along the $x$ (a) and $y$ (b) directions, respectively. (c)-(d) Wavefunctions of the edge states at $k_{x/y} = \pi/a$ and the edge states at frequencies $f_3$ and $f_4$. For $k_{x/y} = \pi/a$, the edge states are either even parity or odd parity (the dashed lines indicate the mirror plane). For the edge states at frequencies $f_3$ and $f_4$, the edge states are either pseudo-spin up or pseudo-spin down. The yellow arrows in the acoustic pressure phase profiles, arg $(P)$, indicate the winding directions of the phase vortices (the yellow dots indicate the vortex centers). The green arrows indicate the momentum ("Poynting vector") distributions of the acoustic edge states. (e) Simulated spectrum for the finite structure (depicted in the inset) with both edge and corner boundaries between the shrunken SC with $\theta = -35°$ and the shrunken SC with $\theta = 60°$. The spectrum of the four degenerate corner states is indicated by the red dots. (f) Acoustic wavefunction (i.e., acoustic pressure profile) of one corner state. The green dashed lines indicate the edge boundaries.

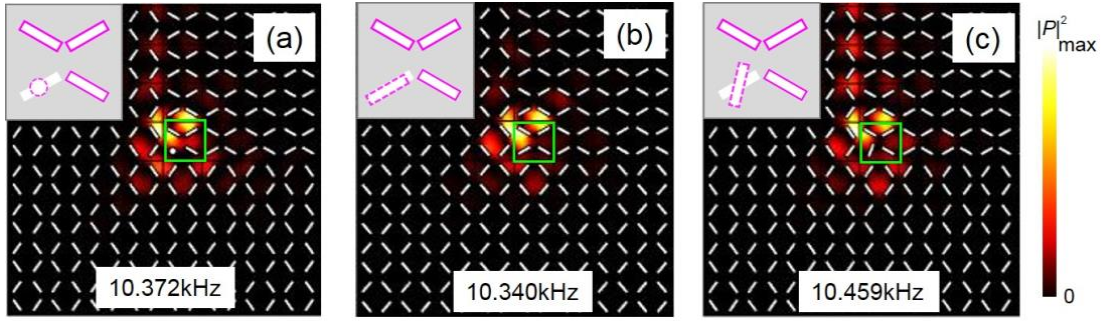

Fig. 9. Robustness of the corner state between the SC with $\theta = -40°$ and the SC with $\theta = 60°$ against deformations. The wavefunctions (hot color) and frequencies (inset) of the corner state when the epoxy block (dashed) in the unit-cell near the corner is replaced by (a) an epoxy dot, (b) a shifted epoxy block, and (c) a rotated epoxy block (rotation angle 20°). The inset at the upper-left show the structure of the deformed unit-cell (magenta) and the undeformed unit-cell (white).

### B. Edge topological transitions in shrunken sonic crystals

We now study the topological transitions at the edges of the shrunken SCs, which reveal the main differences between the shrunken SCs and the original SCs. In Fig. 10(a), we present the evolution of the even and odd modes of the edge states along the $x$ direction at the $k_x = \frac{\pi}{a}$ with the rotation angle $\theta_2$ of the TCI$_\beta$, when the TCI$_\alpha$ has the rotation angle $\theta_1 = -35°$. Since the bulk band gap of the TCI$_\beta$ is finite only for $5° < \theta_2 < 85°$, we study the evolution of the edge states in this region. It is seen from the figure that there is a topological transition in the edge states along the $x$ direction. This transition is signaled by the edge band gap closing at $\theta_2 = 35° = -\theta_1$. Interestingly, although the unit-cells of the SCs have no glide symmetry, the edge boundary along the $x$ direction can have the glide symmetry $G_{edge,x}$, as shown in Fig. 10(b). The emergent glide symmetry $G_{edge,x}$ at the edge along the $x$ direction when $\theta_2 = -\theta_1$ leads to gap-closing edge states as shown in Fig. 10(c). However, the edge boundary along the $y$ direction does not have the glide symmetry [see Fig. 10(e)]. As a consequence, the edge states along the $y$ direction do not experience such a gap closing [see Fig. 10(f)]. As a consequence, there is no topological transition for the $y$ edge at $\theta_2 = -\theta_1$. In fact, the Dirac mass for the $y$ edge, $m_y$, remains positive in the whole region of $5° < \theta_2 < 85°$ [Fig. 10(d)], which is quite different from the situations in the original SCs.

Since the edge Dirac mass, $m_y$, remains positive in the whole region of $5° < \theta_2 < 85°$, the corner states can emerge only in the region when $m_x < 0$. In addition, the $x$ and $y$ edges must have a common spectral gap to stabilize the corner states. We label the region that satisfies these conditions by the dashed box in Figs. 10(a) and 10(d). We indeed found topological corner states in this region labeled by the dashed box.

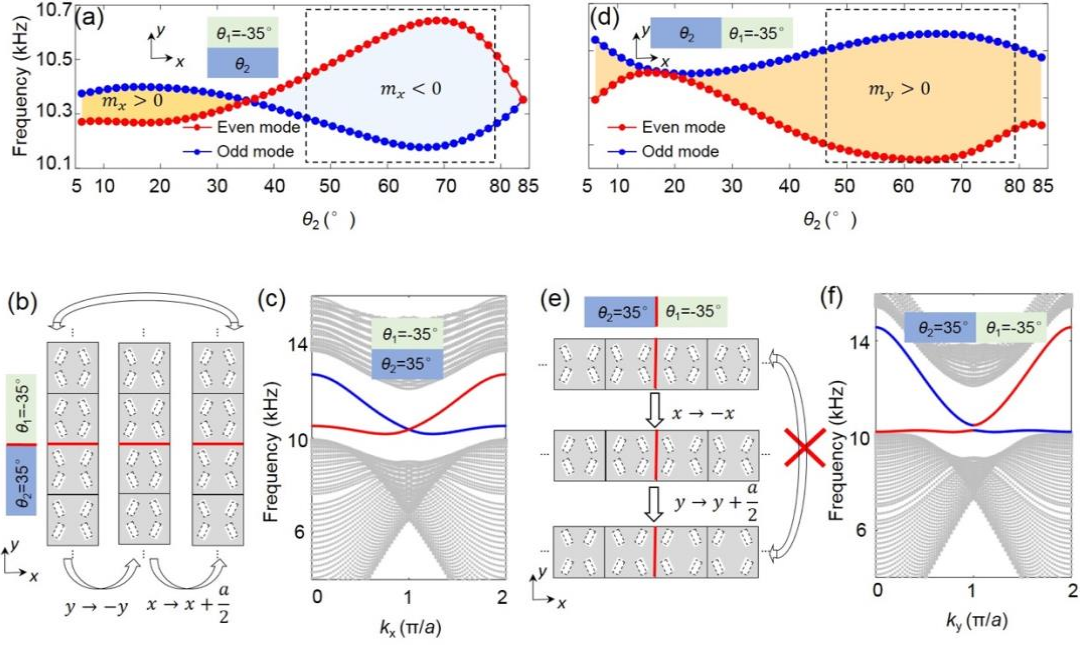

FIG. 10. Topological transitions in a hierarchy of dimensions in the shrunken SCs when $\theta_1 = -35°$. (a) Topological "phase diagram" for the edge states propagating along the $x$ direction as represented by the frequencies of the odd and even modes at $k_x = \frac{\pi}{a}$. The transitions occur at the gap closing points when the odd and even modes become degenerate. (b) Illustration of the glide symmetry at the edge boundary along the $x$ direction for the shrunken SCs when $\theta_2 = -\theta_1$. (c) Gap-closing edge states along the $x$ direction for the case with $\theta_2 = -\theta_1 = 35°$. (d) Topological "phase diagram" for the edge states propagating along the $y$ direction for the shrunken SCs as represented by the frequencies of the odd and even modes at $k_y = \frac{\pi}{a}$. (e) Illustration of the absence of the glide symmetry at the edge boundary along the $y$ direction when $\theta_2 = -\theta_1$. (c) Gapped edge states along the $y$ direction for the case with $\theta_2 = -\theta_1 = 35°$. In (b) and (e), the white blocks represent the location of the epoxy blocks in the original SC, while the dashed blocks represent the location of the epoxy blocks in the shrunken SC.

To study the bulk, edge and corner states versus the rotation angle $\theta_2$ of the shrunken SCs, we calculate the density of states for the bulk, edge and corner regions. The results presented in Fig. 11 confirm our conjecture: the corner resonances appear only in the region labeled by the dashed boxes in Figs. 10(a) and 10(d), i.e., when (i) the edge Dirac masses, $m_x$ and $m_y$, have opposite signs and (ii) there is common spectral gap for the edge states along the $x$ and $y$ directions. Specifically, in the Fig. 11 these conditions are met when $\theta_2 = 60°$ and $70°$ for $\theta_1 = -35°$.

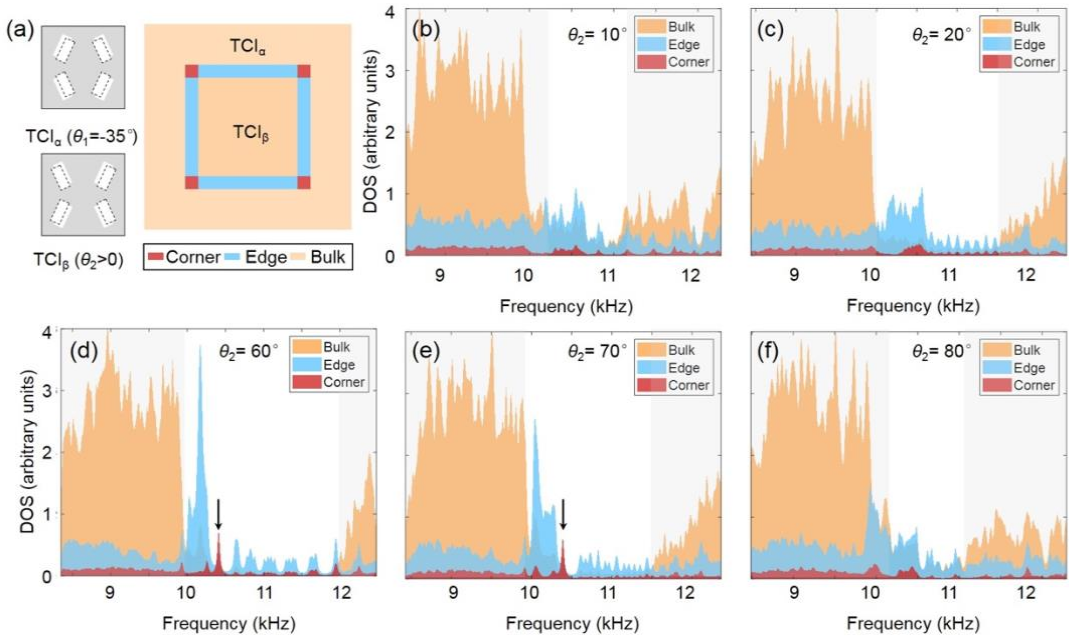

FIG. 11. (a) Illustration of the structure with edge and corner boundaries between the shrunken SC with $\theta_1 = -35°$ and the shrunken SC with $0° < \theta_2 < 90°$. Different colors represent different regions for the calculation of local density of states (DOS). The DOS for the corner, edge and bulk are obtained by integration of local DOS over the corner, edge and bulk regions, respectively. There are four corner (edge) regions. Each corner region contains $2 \times 2$ unit-cells, while each edge region contain $2 \times 10$ unit-cells. In the insets at the left-side, the white blocks represent the location of the epoxy blocks in the original SC, while the dashed blocks represent the location of the epoxy blocks in the shrunken SC. (b)-(f) DOS for the corner, edge and bulk regions for $\theta_2 = 10°, 20°, 60°, 70°$ and $80°$, respectively. The corner resonances are indicated by the black arrows.

## VI. CONCLUSIONS AND OUTLOOK

We study the higher-order topology in a type of $C_2$-symmetric SCs and analyze the underlying physics in the bulk and edge states as well as the topological corners observed in experiments in Ref. [58]. We show the connection between the bulk topology, the gapped edge states and the topological corner states using the effective Hamiltonian theory, the edge polarization theory, and symmetry-representations of the bulk and edge bands. The bulk-edge-corner correspondence is revealed through the topological transitions in both the bulk and the edges. These topological transitions in 2D and 1D indicate rich physics in higher-order topological systems. The multidimensional topological transitions also reveal the essence of higher-order topology: the connection and dependence of topological phenomena in multiple dimensions in a single physical system. The corner states emerge when the edges along the *x* and *y* directions have opposite signs of Dirac mass, or in other words, when they have distinct Zak phases. The emergent corner states demonstrate robustness against deformations even when such deformations are close to the corners. We show that by controlling the rotation angles, the SCs can be highly-tunable systems for the study of higher-order topology. We also study the higher-order topological physics in shrunken SCs where similar but richer bulk and edge topological phenomena are found.

## ACKNOWLEDGEMENTS

Z.X., Z.-K.L., and J.-H.J. acknowledge support from the National Natural Science Foundation of China (Grant No. 11675116), the Jiangsu distinguished professor funding and a project funded by the Priority Academic Program Development of Jiangsu Higher Education Institutions (PAPD). X.J.Z., M.-H.L. and Y.-F.C. acknowledge support from the National Key R&D Program of China (2017YFA0303702 and 2018YFA0306200) and the National Natural Science Foundation of China (NSFC Grants Nos. 11625418, 11890700 and 51732006).